\documentclass[]{pasj01}

\Received{}
\Accepted{}
 
\begin{document} 

\title{ 
\LETTERLABEL 
First detection of PSR B1259--63/LS 2883 in the Millimeter and Submillimeter
Wavelengths with ALMA}

\author{Yutaka \textsc{Fujita}\altaffilmark{1}$^{*}$}%
\altaffiltext{1}{Department of Earth and Space Science, Graduate
	School of Science, Osaka University, Toyonaka, Osaka 560-0043,
	Japan}
\email{fujita@astro-osaka.jp}

\author{Akiko \textsc{Kawachi}\altaffilmark{2}}
\altaffiltext{2}{Department of Physics, School of Science, Tokai
University, Kitakaname, Hiratsuka, Kanagawa 259-1292, Japan}

\author{Takuya \textsc{Akahori}\altaffilmark{3}}
\altaffiltext{3}{National Astronomical Observatory of Japan, 2-21-1 Osawa, Mitaka, Tokyo 181-8588, Japan}

\author{Hiroshi \textsc{Nagai}\altaffilmark{3}}

\author{Masaki \textsc{Yamaguchi}\altaffilmark{4}}
\altaffiltext{4}{Department of Physics, Faculty of Science and
Engineering, Konan University 8-9-1 Okamoto, Kobe, Hyogo 658-8501,
Japan}


\maketitle

\begin{abstract}
We report Atacama Large Millimeter/submillimeter Array (ALMA)
observations of the binary system containing the pulsar PSR~B1259--63
orbiting around a Be star LS~2883 after the 2017 periastron passage. We
detected radio continuum emission from the binary system in the
millimeter/submillimeter wavelengths for the first time. At Band~3
(97~GHz), the flux 84~days after the periastron is almost the same as
that 71~days after the periastron. Although the binary system showed
intense GeV gamma-ray flares during our observations, the Band~3 flux
did not indicate any time correlation with them. The Band~3 fluxes are
consistent with an extrapolation of the radio spectrum at lower
frequencies. Assuming that it is synchrotron emission, we constrain
magnetic fields ($\lesssim 0.6$~G) and the high-energy cutoff of the
electrons ($\gamma \gtrsim 360$). The flux at Band~7 (343~GHz) 69~days
after the periastron shows a significant excess from the extrapolation
of the radio spectrum at lower frequencies. The flux may be associated
with the circumstellar disk around the Be star. We also present the
results of Australian Telescope Compact Array (ATCA) observations at
94~GHz for the 2014 periastron passage, which show that the radio
spectrum was relatively soft when the pulsar passed the disk.
\end{abstract}

\section{Introduction}

PSR~B1259--63 is a $48$~ms period rotation-powered pulsar in a highly
eccentric ($e=0.87$), 3.4~year orbit with a Be star LS~2883
\citep{1992ApJ...387L..37J}.  The distance to the
binary is inferred to be 2.6\,kpc \citep{2018MNRAS.479.4849M}.  Be
stars are known to have anisotropic stellar winds and
an equatorial disk with enhanced mass outflow.  Pulsar-timing data
indicates that the circumstellar disk is inclined to the plane of the binary orbit
\citep{{1995MNRAS.275..381M}}.
The pulsar's orbital radius at periastron is $\sim 21\: R_*$, where $R_*\sim 9.2\: R_\odot$ is the radius of the Be star LS~2883
\citep{2011ApJ...723L..11N}.  Because of the highly elliptical orbit,
PSR B1259--63 is expected to interact with the equatorial disk of the Be
star near the periastron
\citep{2011PASJ...63..893O,2012ApJ...750...70T}.

The binary has been observed in multi-wavelengths. A double peaked
non-thermal and unpulsed radio outburst takes place around a periastron
passage, each of which reflects the disk crossing of the pulsar
\citep{2002MNRAS.336.1201C,2005MNRAS.358.1069J}. The radio emission
seems to be synchrotron emission from electrons accelerated at the shock
produced through the interaction between the pulsar wind and the disk or
the stellar outflow.  X-ray light curves also show the double-peak
structure; this emission may be synchrotron emission or inverse Compton
(IC) scattering
\citep{2006MNRAS.367.1201C,2009MNRAS.397.2123C,2009ApJ...698..911U,2011MNRAS.417..532P,2015MNRAS.454.1358C}.

Unpulsed TeV gamma-ray emission has been detected with High Energy
Stereoscopic System (HESS) close to the periastron passage
\citep{2005A&A...442....1A,2009A&A...507..389A}. If the emission is IC
scattering, the intense photon fields provided by the companion star and
the disk serve as a target for the production of the high energy
gamma-rays
\citep{1996A&AS..120C.221T,1999APh....10...31K,2011MNRAS.416.1067K},
although hadronic mechanisms may be possible
\citep{2004ApJ...607..949K,2007Ap&SS.309..253N}. In the GeV band, flares
have been observed with the \textit{Fermi Gamma-ray Space Telescope}
after the periastron passage and after the passage of the dense
equatorial disk of the Be star
\citep{2011ApJ...736L..11A,2015ApJ...811...68C,2018ApJ...862..165T,2018ApJ...863...27J}. The
flares are apparently triggered by the disruption of the circumstellar
disk \citep{2015MNRAS.454.1358C}.

Despite these multi-wavelength observations, there are yet-to-be
explored bands. While previous radio observations have been limited to
the frequencies of $\nu\lesssim 10$~GHz, no information has been
obtained at $\nu\gtrsim 10$~GHz. Observations in this band could elucidate
a link between the radio and X-ray emissions. For example, if there is a
cutoff in this band, the X-ray emission is likely to be produced by a
mechanism that is different from that of the radio emission. In that
case, IC scattering is a likely solution \citep{2006MNRAS.367.1201C}. If
not, the X-ray emission could be synchrotron emission, which is the same
as the radio emission, although it is not conclusive. Moreover, the
emission from the circumstellar disk around the Be star could be
observed in this band. It is important to reveal this emission because
the radiation field can affect the gamma-ray emission from the binary
through IC scattering
\citep{2011MNRAS.412.1721V,2012MNRAS.426.3135V}. 

In this letter, we report the detection of the binary system
PSR~B1259--63 at 97 and 343~GHz after the 2017 periastron passage with
Atacama Large Millimeter/submillimeter Array (ALMA) for the first
time. We also present Australian Telescope Compact Array (ATCA)
observations at 94~GHz for the 2014 periastron passage.

\section{Observations}
\subsection{ALMA}

ALMA Band~3 ($\lambda=3\:\rm mm$) observations were carried out on 2017
December 2 with forty-two $12\rm\: m$ antennas and December 15 with
forty-five $12\rm\: m$ antennas, and Band~7 ($\lambda=0.9\:\rm mm$)
observation was carried out on 2017 November 30 with forty-seven
$12\rm\: m$ antennas (project code: 2017.1.01188.S).  The data were
taken with time division mode (TDM) centered at 97~GHz for Band~3 and
343~GHz for Band~7.  The total on-source time is about 5~minutes for all
three observations.  The data were processed with the CASA/ALMA Pipeline
(Pipeline-CASA51-P2B) in a standard manner.  Bandpass characteristics
were corrected in both amplitude and phase using bandpass calibrator
J1647-5848 and J1337-1257 in Band~3 and Band~7, respectively.  Gain
calibration was done using phase calibrator J1322-6352 and J1308-6707
in Band~3 and Band~7, respectively, with the aid of phase correction
with the water vapor radiometer.  Amplitude scaling was derived with the
bandpass calibrator.

Images were created by combining all spectral windows in each band.
Deconvolution was done using the CASA task {\tt tclean} with the Briggs
weight of robust parameter of 0.5.  Table \ref{tab:ImageSummary}
summarizes the image rms and angular resolution. We refer to the
periastron passage time as $t_{\rm p}$ regardless of periastron
passages.  We detected a compact, single component in both Band~3 and~7.
The flux was evaluated as the total flux density derived by a Gaussian
model fitting.

We note that the ALMA fluxes include the pulsed component of the
pulsar. However, the observations at lower frequencies ($\sim$~a few
GHz) show that the flux of the pulsed component is a few mJy and the
component has a spectrum approximately given by $\propto \nu^{-1}$
\citep{2005MNRAS.358.1069J}. Thus, it can safely be ignored at the ALMA
bands.

\begin{table*}
  \tbl{Angular resolution, image rms, and observed fluxes for the ALMA
  observations in 2017}{
  \begin{tabular}{cccccc}
 \hline
Band & Date & Day & Beam Shape & Image RMS  & Observed Flux  \\ 
     & & (from $t_{\rm p}$) &  & ($\mu$Jy~beam$^{-1}$) & (mJy)  \\ \hline   
 3 (97~GHz) & Dec 2 & +71 & $0.35"\times0.21"$ at $78^\circ$ & 41 & $1.1\pm 0.1$ \\
 3 (97~GHz) & Dec 15 & +84 & $0.42"\times0.36"$ at $-52^\circ$ & 36 & $0.97\pm 0.09$\\
 7 (343~GHz) & Nov 30 & +69 & $0.056"\times0.043"$ at $-8^\circ$ & 87 & $2.3\pm 0.4$\\ \hline
  \end{tabular}}\label{tab:ImageSummary}
\end{table*}

\begin{table*}
\tbl{The calibrators and observed fluxes for the ATCA observations in 2014}{
  \begin{tabular}{ccccccc}
   \hline
  Date & Day & Observing Time & \multicolumn{3}{c}{Calibrators}  & Residual RMS ($1\:\sigma$)  \\ \cline{4-6}
   & (from $t_{\rm p}$) &  (min) & Bandpass & Flux & Gain/Phase  & (mJy)  \\
    \hline 
    Apr 4 & $-$29.9  & 20 & PKS~1253$-$055 & Mars  & PKS~1305$-$668 & 3.58 \\
    Apr 6 & $-$27.8  & 286 & PKS~0537$-$441 & Jupiter  & PKS~1305$-$668 & 0.853 \\
    May 19 & $+$15.4 & 82  & PKS~1921$-$293 & Uranus  & PKS~1305$-$668 & 1.550 \\
    May 27 & $+$23.3 & 68 & PKS~1253$-$055 & Uranus  & PMN~J1326$-$5256 & 0.704 \\
    Jun 15 & $+$42.2 & 74 & PKS~1921$-$293 & Saturn  & PMN~J1326$-$5256 & 0.466 \\
    Jun 29 & $+$56.1  & 71 & PKS~1253$-$055 & Mars  & PMN~J1326$-$5256 & 0.139 \\
    Sep 26 & $+$145.6 & 58 & PKS~0537$-$441 & Jupiter  & PMN~J1326$-$5256 & 1.43 \\
\hline
\end{tabular}}
\label{tab:ATCAobs}
\end{table*}

\subsection{ATCA}

The observations at $3\rm\: mm$ wavelength (94~GHz) with ATCA were also
carried out around and after the 2014 periastron passage of the
binary. The ATCA consists of six $22\rm\: m$ radio antennas of which
five are available for 3~mm observations.  The Compact Array Broadband
Backend (CABB) provides observations with two 2048~MHz intermediate
frequencies (IF) bands which are centered at 93.504~GHz and 95.552~GHz.
The target was observed with several sets of $10\rm\: minute$ on-source
observations in between the observations of the gain/phase
calibrator. The calibrators for the bandpass, flux density and
gain/phase of each observation are summarized in
Table~\ref{tab:ATCAobs}.  
The pulsar-binning mode was not employed in the observations based 
 on the similar pulse flux estimation in the previous ALMA section.
The data was processed using the data reduction
package MIRIAD \citep{1995ASPC...77..433S}.  A Gaussian fit was
performed on the Stokes {\it I} images with the MIRIAD task {\tt imfit},
and when no detection was confirmed, the residual rms error over the
image (table~\ref{tab:ATCAobs}) is used to obtain the upper limit
(figure~\ref{fig:ATCA}). The inverted images are produced to minimize
the noise level for better sensitivity.

\section{Results and Discussion}

\begin{figure}
 \begin{center}
  \includegraphics[width=8cm]{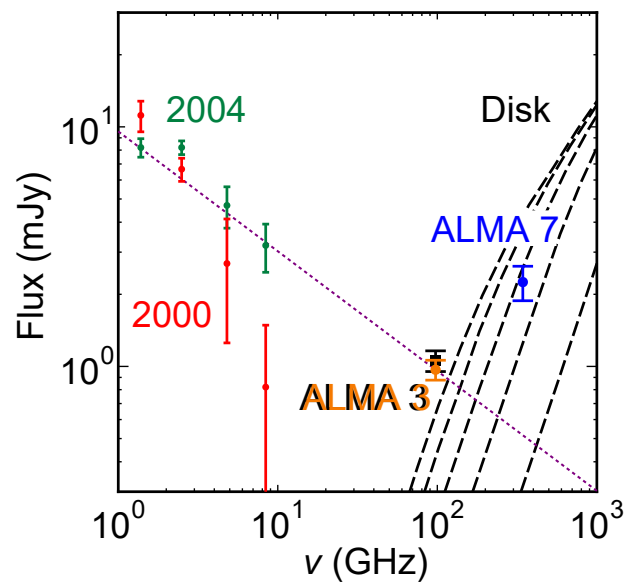}
 \end{center}
\caption{Radio fluxes of PSR~B1259--63. Those obtained with ALMA are
shown by the filled black square (Band~3 or 97~GHz at $t_{\rm
p}+71$~days), the filled orange circle (Band~3 at $t_{\rm p}+84$~days),
and the filled blue circle (Band~7 or 343~GHz at $t_{\rm
p}+69$~days). Unpulsed fluxes obtained with ATCA at $t_{\rm p}+63.4$~days
at the 2000~passage are shown by red dots \citep{2002MNRAS.336.1201C}
and those at $t_{\rm p}+64.2$~days at the 2004~passage are shown by
green dots \citep{2005MNRAS.358.1069J}. The dotted
purple line is a spectrum represented by $S_\nu \propto \nu^{-0.5}$. The
normalization is set so that the line passes the second ALMA Band 3
observation (filled orange circle). Dashed black lines show the
infrared emission from the Be star's circumstellar disk for different
disk sizes predicted by \citet{2011MNRAS.412.1721V}. The disk sizes are
(from the bottom to the top) 10, 20, 30, 40 and 50~$R_*$. }\label{fig:spec}
\end{figure}

\begin{figure}
 \begin{center}
  \includegraphics[width=8cm]{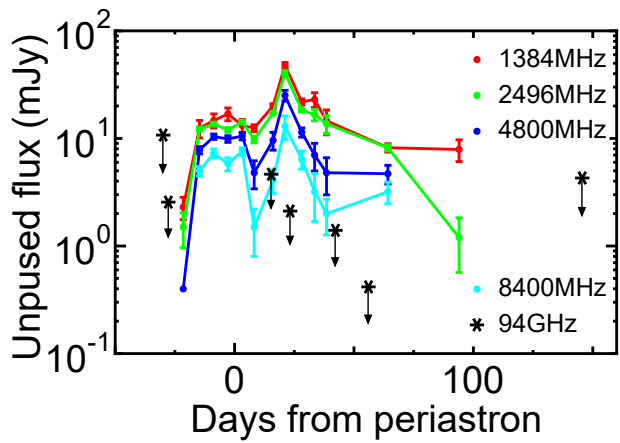}
 \end{center}
\caption{ATCA upper limits (3~$\sigma$) at 94~GHz for the 2014 periastron
passage are shown by stars. Lines are light curves of the unpulsed
emission for the 2004 periastron passages \citep{2005MNRAS.358.1069J}. Each
line displays four different frequencies, 1384, 2496, 4800 and 8640~MHz,
in order from top down.\label{fig:ATCA}}
\end{figure}

\begin{figure}
 \begin{center}
  \includegraphics[width=8cm]{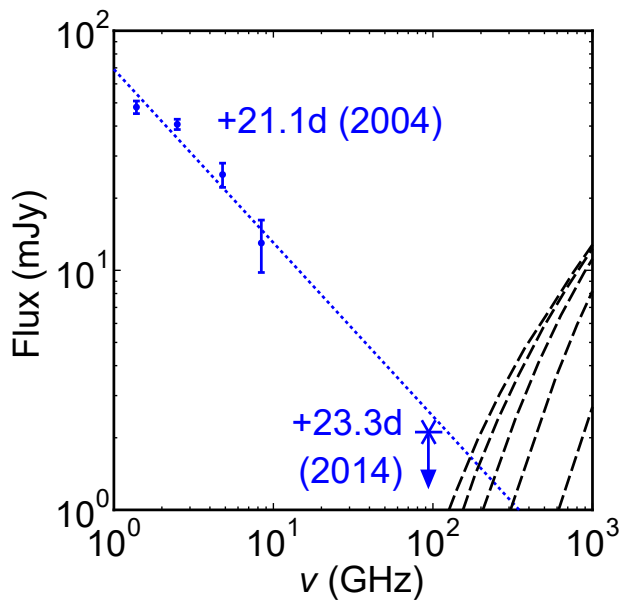}
 \end{center}
\caption{Stars are ATCA upper limits (3~$\sigma$) at 94~GHz for the 2014
periastron passage at $t_{\rm p}+23.3$ days. Dots are fluxes for the 2004
periastron passage at $t_{\rm p}+21.1$ days
\citep{2005MNRAS.358.1069J}. Dotted line is a power-law fit for the
data for the 2004 periastron passage ($\propto \nu^{-0.72}$). Dashed black lines show the
infrared emission from the Be star's circumstellar disk, and they are the same as those in figure~\ref{fig:ATCA}. \label{fig:ATCA_spec}}
\end{figure}

The most recent periastron passage of the PSR B1259--63/LS~2883 system
occurred on MJD 58018.143 (UTC 2017 September 22 03:25:55.2). In
table~\ref{tab:ImageSummary} and figure~\ref{fig:spec}, we show the
radio fluxes of PSR~B1259--63 obtained with ALMA. At 97~GHz, we observed
the object twice ($t_{\rm p}+71$ and~$t_{\rm p}+84$ days) and found that
their fluxes are identical within the errors. For a comparison, we show
unpulsed (total minus pulsed) radio fluxes at 1384, 2496, 4800, and
8400~MHz obtained with ATCA at similar post-periastron days at the 2000
and 2004 periastron passages. Unfortunately, observational results for
the 2007--2017 periastron passages at those low frequencies have not
been published. However, \citet{2005MNRAS.358.1069J} indicated that the
light curve of the unpulsed radio emission at 1384~MHz is broadly
similar from one periastron to the next. The spectrum at the 2000
periastron passage is steeper than that at the 2004 periastron passage
(figure~\ref{fig:spec}). While for the former the fluxes at 4.8 and
8.4~GHz may be affected by phase-calibration errors
\citep{2002MNRAS.336.1201C}, we cannot deny the possibility that the
spectrum at those and higher frequencies is highly time dependent.  The
thin dotted purple line represents a spectrum of $S_\nu \propto
\nu^{-0.5}$. It shows that the ALMA Band~3 fluxes can be regarded as an
extrapolation of the ATCA fluxes for the 2004 periastron passage. The
power-law spectrum with a negative index implies that the radio emission
is a synchrotron emission.

PSR~B1259--63 is known to become active after a periastron passage in
the GeV gamma-ray band. For the 2017 periastron passage, the gamma-ray
activities were more intense than the previous ones and displayed
variability of short timescales ($\sim 1.5$~minutes). These activities
were observed from $\sim t_{\rm p}+40$~days to $\sim t_{\rm p}+75$ days
\citep{2018ApJ...862..165T,2018ApJ...863...27J}. Thus, while our first
Band~3 observation ($t_{\rm p}+71$~days) was performed during the active
period, the second observation ($t_{\rm p}+84$~days) was done outside
the period. The stable radio flux between the two observations suggests
that the origin of the radio synchrotron emission is different from that
of the gamma-ray emission.
 
Figure~\ref{fig:spec} indicates that the flux at the ALMA Band~7
(343~GHz) at $t_{\rm p}+69$~days is clearly above the extrapolation from
the fluxes at low frequencies (thin dotted purple line). This suggests
that the flux is not attributed to the synchrotron emission. The most
plausible explanation is that the emission comes from the circumstellar
disk around the Be star, which supplies seed photons for gamma-ray
production through IC scattering
\citep{2012ApJ...752L..17K,2012MNRAS.426.3135V}. The dashed black lines
in figure~\ref{fig:spec} show a prediction of the disk emission
\citep{2011MNRAS.412.1721V}. The Band~7 flux is consistent with the
spectrum when the disk radius is $\sim 30\: R_*$. This radius is smaller
than the one simply expected from pulsar eclipses ($\sim 50\: R_*$;
\cite{2011MNRAS.412.1721V}). The discrepancy may be caused by disruption
of the disk by the pulsar passage or by model uncertainties. In order to
discriminate the two possibilities, it would be useful to observe the
object at different times before and after the disk passage and see if
the flux of the possible disk component changes.

The Band 3 fluxes are above the predicted disk emission
(figure~\ref{fig:spec}), which also suggests that they are synchrotron
emission.  The extension of the synchrotron emission up to 97~GHz
constrains the magnetic fields of the system ($B$) if electrons are
instantaneously injected. Since the spectrum is presented by a
power-law, electrons that radiate the emissions at $\lesssim 97$~GHz are
not affected by synchrotron loss until at least $\sim t_{\rm
p}+84$~days.  Assuming that synchrotron emission at the frequency of
$\nu$ is produced by electrons with a Lorentz factor of $\gamma$, the
relation between $\nu$ and $\gamma$ is
\begin{equation}
\label{eq:nu}
 \nu\sim 0.29\:\nu_c
\equiv 0.29\frac{3\gamma^2 e B\sin\alpha}{4\pi m_e c}
\:,
\end{equation}
where $e$ is the unit charge, $\alpha$ is the pitch angle of the
electrons, $m_e$ is the electron mass, and $c$ is the light speed
\citep{ryb79}. If electrons with a power-law spectrum are
instantaneously injected at $t=0$, and if they passively evolve being
affected by synchrotron loss, the electrons with Lorentz factors larger
than $\gamma$ will have been removed at $t=T$, where
\begin{equation}
\label{eq:gamma}
 \gamma = \frac{1}{b_1 T}
\end{equation}
\citep{1999ApJ...520..529S}. The factor $b_1$ is related to the
synchrotron loss rate for $\gamma\gg 1$:
\begin{equation}
 -\frac{d\gamma}{dt} 
= \frac{4}{3}\frac{\sigma_{\rm T}}{m_{\rm e} c}\frac{B^2}{8\pi}\gamma^2
\equiv b_1 \gamma^2
\end{equation}
\citep{1999ApJ...520..529S}. By eliminating $\gamma$ from
equations~(\ref{eq:nu}) and~(\ref{eq:gamma}), the magnetic field is
written as
\begin{equation}
\label{eq:B}
 B = 0.6\: (\sin\alpha)^{1/3}\left(\frac{\nu}{97\rm\: GHz}\right)^{-1/3}
\left(\frac{T}{64\rm\: days}\right)^{-2/3}\rm\: G\:.
\end{equation}
Here, we assumed that the electrons were injected when the pulsar passed
through the disk around the Be star at $\sim t_{\rm p}+20$ days
\citep{2005MNRAS.358.1069J}. If the electrons are affected by adiabatic
cooling and the magnetic-field strength is a decreasing function of time
\citep{1997ApJ...477..439T,2002MNRAS.336.1201C}, the field at the
observation ($\sim t_{\rm p}+84$~days) must be weaker. Thus,
equation~(\ref{eq:B}) shows that the magnetic field associated with the
synchrotron emission is $B\lesssim 0.6$~G. This field strength is
roughly consistent with the predicted ones around the shock created
through the interaction of the pulsar wind with the outflow from the Be
star \citep{1997ApJ...477..439T,1999ApJ...514L..39B}, and the values
deduced from the observed radio, X-ray and gamma-ray emissions near the
periastron \citep{2002MNRAS.336.1201C,2005A&A...442....1A}. If we assume
that $B=0.6$~G and $T=64$~days, the Lorentz factor is $\gamma=360$ 
[equation~(\ref{eq:gamma})]. Thus, the cooling cutoff in the electron
energy spectrum should be at $\gamma>360$. If the spectrum further
extends to higher energies, the X-ray emission of the binary can also be
explained by the synchrotron emission.

For the 2014 periastron, which happened on 2014 May~4 ($t_{\rm p}$ =
MJD~56781.42), we obtained only image rms ($1\:\sigma$) at 94~GHz with
ATCA (table~\ref{tab:ATCAobs}), from which we estimate $3\;\sigma$
upper limits of the flux (figure~\ref{fig:ATCA}). For comparison, we
present light curves for the 2004 periastron passage; they have a peak
at $\sim t_{\rm p}+20$~days, which corresponds to the disk passage of
the pulsar \citep{2005MNRAS.358.1069J}.  Figure~\ref{fig:ATCA_spec}
shows the radio spectrum at the disk passage. The upper limit at 94~GHz
is below the extrapolation of lower frequency fluxes ($\propto
\nu^{-0.72}$). This means that the spectrum is relatively soft and is
not hardened during the passage. The relatively small flux at 94~GHz
could constrain theoretical models of particle acceleration through
interaction between the pulsar wind and the disk.  For example, the
acceleration would take some time to form a power-law spectrum up to the
high frequency.

\section{Conclusion}

We have reported the results of ALMA observations of the pulsar-Be star
binary PSR~B1259--63/LS~2883 at 97 and 343~GHz after the 2017 periastron
passage. We detected the object at both frequencies for the first
time. We found that the flux at 97~GHz does not change between our two
observations (interval of 13~days) and it does not have a time
correlation with intense gamma-ray flares. The flux is consistent with
an extrapolation of the spectrum at lower frequencies of $\lesssim
10$~GHz, which means that the radio synchrotron emission extends at
least up to 97~GHz. Assuming that the electrons that are responsible for
the synchrotron emission are injected when the pulsar passes through the
circumstellar disk of the companion star, the magnetic-field strength
and the cooling-cutoff energy of the electrons can be constrained to be
$B\lesssim 0.6$~G and $\gamma\gtrsim 360$, respectively. The flux at
343~GHz is much larger than the extrapolation of the low-frequency
observations. This may indicate that the emission comes from the
circumstellar disk of the Be star. We have also presented ATCA
observations at 94~GHz for the 2014 periastron passage. They showed that
the radio spectrum is not hardened during the disk passage of the
pulsar. Our results could be new information that constrains theoretical
models.

\begin{ack}
This work was supported by MEXT KAKENHI No.~18K03647 (Y.F.) and
JP18K03709 (H.N.). M.Y. was supported by Hayakawa Fund.  We acknowledge
Dr. Jamie Stevens, CSIRO's Senior Systems Scientist at the Australian
Telescope Compact Array for his support. This paper makes use of the
following ALMA data: ADS/JAO.ALMA\#2017.1.01188.S. ALMA is a partnership
of ESO (representing its member states), NSF (USA) and NINS (Japan),
together with NRC (Canada), MOST and ASIAA (Taiwan), and KASI (Republic
of Korea), in cooperation with the Republic of Chile. The Joint ALMA
Observatory is operated by ESO, AUI/NRAO and NAOJ.
\end{ack}


\end{document}